\documentclass[journal=nalefd,manuscript=article]{achemso}
\usepackage[version=3]{mhchem} 
\usepackage{bm}
\usepackage{color}
\usepackage{xcolor} 

\author{Pietro Tierno}
\affiliation{$^\dag$ Departament de F\'isica de la Mat\`eria Condensada, Universitat de Barcelona, Av. Diagonal 647, 08028 Barcelona, Spain}
\affiliation{$^\ddag$ Institut de Nanoci\`encia i Nanotecnologia,
Universitat de Barcelona, 08028, Barcelona, Spain}
\email{ptierno@ub.edu}
\author{Tom H. Johansen}
\affiliation{$^{\P}$ Department of Physics, The University of Oslo,
P.O. Box 1048 Blindern, 0316 Oslo, Norway\,}
\affiliation{$^\S$ Institute for Superconducting and Electronic
Materials, University of Wollongong, Australia\,}
\author{Jose M. Sancho}
\affiliation{$^\dag$ Departament de F\'isica de la Mat\`eria Condensada, Universitat de Barcelona,
Av. Diagonal 647, 08028 Barcelona, Spain}
\affiliation{$^\ddag$ Institut de Nanoci\`encia i Nanotecnologia,
Universitat de Barcelona, 08028, Barcelona, Spain\,}
\title{A Tuneable Magnetic Domain Wall Conduit
Regulating Nanoparticle Diffusion}
\begin{document}

\begin{abstract}
We demonstrate a general and robust method to confine
on a plane strongly diffusing submicrometer particles
in water by using size tunable
magnetic channels. These virtual conduits are realized with
pairs of movable Bloch walls (BWs)
located within an
epitaxially grown ferrite garnet film.
We show that, once inside the magnetic conduit,
the particles
experience an effective
local parabolic potential
in the transverse direction, while freely diffusing
along the conduit.
The stiffness of the magnetic potential
is determined as a function of field amplitude
which varies the width of the magnetic channel,
and precise control of
the degree of confinement is demonstrated
by tuning the applied field.
The magnetic conduit is then used to realize single files of
non-passing particles and to induce periodic
condensation of an ensemble of particles
into parallel stripes in a
completely controllable and reversible manner.
\end{abstract}
The ability to confine nanoscale
particles in a fluid medium
is of critical importance in a number of
fields related to microfluidics,~\cite{Val2012}
optics~\cite{Gri2003} and biophysics.~\cite{Col2005}
Optical tweezers based on
focusing an intense laser beam to a
diffraction limited spot, demonstrated
in the past trapping,~\cite{Ash1970} rotation,~\cite{Che02}
and transport~\cite{Fau1995}
of microscopic particles in water.
Confinement of multiple microspheres in two and three dimensions
has also been obtained with
more sophisticated techniques such as
fast scanning beams~\cite{Hoo2002}
and holographic optical tweezers.~\cite{Pad2011}
However,
reducing the size of the
colloidal particles to the nanoscale
makes
stable trapping via optical means difficult,
due to both the increase of thermal fluctuations,~\cite{Mar2013}
and the need for large and localized optical force gradients.~\cite{Leh2015}
While individual
gold~\cite{Svo94,Han37} and silver~\cite{Bos08} nanoparticles
have been
immobilized,
confinement of the particles over an extended
area, much larger than
a single trapping spot,
remains elusive.\\
Recent experiments using
nanostructured slits made by
electron-beam lithography,
propose a different, non optical, approach
to levitate and confine single
nanoparticles along predefined tracks.~\cite{Kri2010}
These topographic reliefs induce a spatial
modulation of the
electrostatic potential, which attracts
charged nanoparticles
located at a close distance.
However, the dynamics of colloidal particles moving near
lithographic structures
can be affected by the presence of the hard-walls
due to hydrodynamic~\cite{Cui2004} or steric~\cite{Bev2000} interactions,
and be rather sensitive to the
ionic concentration of the
dispersing medium.\\
Magnetic manipulation of
polarizable particles is an alternative approach to control
colloidal matter
not relying
on the particle surface charge or
the dielectric constant,
but rather on its magnetic content.
Magnetic nanoparticles can find applications in
disparate fields, from
spintronics,~\cite{Bha16} to
biomedical research,~\cite{Abo12}
hyperthermia~\cite{Ban13}, or as active propellers.~\cite{Ale16}
While several approaches demonstrated forms of controlled 
trapping of microscale
colloids above magnetic tracks,~\cite{Yel05,Gun05,Don10,Ehr11,Rap12,Dem13}
to our knowledge,
only Ommering and coworkers~\cite{Kim2006}
were able to confine 
a single magnetic nanoparticle
on a microfluidic chip.\\
Here we demonstrate a different
approach which allows to obtain stable
confinement in a fluidic environment of magnetic particles on a plane.
Our technique is based on the use of
remotely controllable
domain walls in a ferromagnetic
structured substrate, 
namely a 
uniaxial ferrite garnet
film (FGF).
The research on domain wall dynamics in magnetic
systems has recently
drawn much interest because of
new applications in spintronics,~\cite{Sev2014} 
logic devices~\cite{Fra2012,Phu2015},
nanowires~\cite{Gu2007,Xin2011}
and ultracold atoms.~\cite{Wes2012}
Our magnetic conduit
makes use of Bloch walls (BWs),
i.e. transition regions where
the magnetization
rotates by 180 degrees through the wall
separating opposite
magnetized domains.
The BWs in the used FGF can be displaced by applying
relatively small external field,
and they generate strong local stray fields
capable to trap and confine magnetic microspheres.~\cite{Hel03}
In this work we show that these highly  functional features can be
extended to nanoscale system,
by tailoring the magnetic attraction of the film.
In particular, we create a magnetic conduit for submicrometer particle motion
using parallel planar BWs which provide
locally a tuneable parabolic-like potential.
These conduits are shown to be able
to regulate the diffusive motion
of an ensemble of particles,
realizing
e.g., single-file conduits for particle motion,
where mutual passage is excluded.
By applying an oscillating magnetic field,
we vary the energy landscape
and demonstrate periodic particle
assembly and condensation
into parallel bands.\\
The magnetic conduits
are realized on the free surface of a FGF
of composition Y$_{2.5}$Bi$_{0.5}$Fe$_{5-q}$Ga$_{q}$O$_{12}$
($q=0.5-1$),
thickness $\sim 4\mu m$
and having a saturation magnetization (the FGF)
of $M_s=2.7 \, {\rm kA/m}$.
The FGF is grown
by dipping liquid phase epitaxy
on a $\langle 111 \rangle$ oriented single crystal
gadolinium  gallium  garnet (Gd$_3$Ga$_5$O$_{12}$)
substrate.~\cite{Tie2009}
Fig.1(a) shows
a typical transparent FGF sample used during the experiments.
Due to the polar Faraday effect,
the magnetic domains
in the
FGF can be visualized
via polarized light microscopy.
As shown in Fig.1(b), under zero applied field
the FGF is characterized by
up- and
down-magnetized domains which
appear as stripes of
black and white colors,
repeating with
a spatial periodicity $\lambda=6.8 \, {\rm \mu m}$.
The BWs on the other hand, given their small width ($\sim 20 nm$),
cannot be resolved by optical microscopy.
The size of the magnetized domains can be easily
manipulated via an external field.
A magnetic field perpendicular to the
film $H_z$, increases the width of the domains with
magnetization
parallel to the field ($\lambda_{+}$) and decreases the width of the antiparallel
ones ($\lambda_{-}$). As shown in Fig.1(c), for $H_z<3 \, {\rm kA /m}$
both, $\lambda_{+}$ and $\lambda_{-}$ vary linearly with $H_z$,
and the total width $\lambda=\lambda_{+}+\lambda_{-}$,
remains
essentially constant.
In contrast, for higher field $\lambda_{+}$ and $\lambda$ diverge,
showing large and irreversible deformations of the pattern. Beyond a critical value,
the film becomes one single magnetized domain.~\cite{Joh13}\\
We use three types of superparamagnetic polystyrene particles with
diameters
$d_p=540{\rm nm}$, $360{\rm nm}$ and $270 {\rm nm}$,
COOH surface groups and $\sim 40 \% \, {\rm wt.}$
iron oxide content (Microparticles GmbH).
Each type of particle is diluted in high deionized water
and deposited above the
FGF, where they sediment
due to the magnetic attraction of the BWs.
Once at the surface of the FGF,
and in absence of any coating
of the film,
the strong attraction of the BWs
rapidly immobilize the submicrometer particles,
which show weak thermal fluctuations,
see
MovieS1 in the
Supporting Information.
In order to
decrease the strong magnetic attraction,
we coat the FGF film
with a $h=1.2 {\rm \mu m}$ thick
layer of a photoresist (AZ-1512 Microchem, Newton, MA),
i.e. a light curable polymer matrix,
using spin coating at $3000$ rpm for $30s$
(Spinner Ws-650Sz, Laurell) 
and $5s$ of UV photo-crosslink (Mask Aligner MJB4, SUSS Microtec).
Since the stray field of the
FGF decreases exponentially from the surface,
after coating the FGF with the polymer layer, 
the submicrometer particles display
a stronger diffusive dynamics,
but still remain two dimensionally (2d) confined above
the FGF with negligible
out of plane motion.
Increasing the particle elevation from the film
not only reduces the
attractive force toward the substrate, but
also modifies the shape of
the potential, as schematically illustrated in
Fig.1(d).
Under a perpendicular field $H_z$,
the magnetic potential at an elevation 
$z_1=d_p/2$ has minima located at the BWs
and maxima between them.
In contrast, at an elevation $z_2=d_p/2+h$,
the potential minima (maxima) are now located at the center of the
large (small) domains, and feature a local parabolic-like shape.\\
We can quantify the magnetostatic potential of the FGF,
by calculating the energy of interaction  $U_{m}$
of one  paramagnetic particle
with the magnetic potential
at a given elevation $z$.
When subjected to a total magnetic field $\bm{H}_{tot}$,
a paramagnetic nanoparticle
acquires a dipole moment $\bm{m}=V\chi \bm{H}_{tot}$,
pointing along the field direction.
Here $V=\pi d_p^3/6$ is the particle volume,
$\chi \sim 2$ the effective volume susceptibility~\cite{Kim2006}
and $\bm{H}_{tot}=\bm{H}_{a}+\bm{H}_{s}$
the sum of the applied
field $\bm{H}_{a}$ and the stray field $\bm{H}_{s}$
of the FGF.~\cite{Str2013} 
The interaction energy
of the induced moment is thus given by,
$U_m= -\frac{1}{2}\mu_w \bm{m}\cdot \bm{H}_{tot}$
where $\mu_w\sim\mu_0 = 4 \pi \cdot 10^{-7} {\rm H /m}$
is the magnetic permeability of the medium (water).
Figs.2(a,c) show the normalized energy landscape
$U_m/k_B T$ numerically evaluated
for a $360nm$
particle (elevation $z=h + d_p/2=0.20 \lambda$)
in absence of an external field, Fig.2(a), and for
an applied field of amplitude $H_z=620 \, {\rm A/m}$, Fig.2(c).
Here $k_B$ is the Boltzmann
constant and $T=293.15 {\rm K}$
the experimental temperature.
In absence of the applied field, the
FGF surface displays a sinusoidal-like
potential at the particle elevation,
blue curve in Fig.2(a), 
with
several small wells which can be easily crossed
by the fluctuating particle.
Effectively, the corresponding particle trajectory
in Fig.2(b)
shows the submicrometer particle
performing simple Brownian diffusion and
passing the BWs without any perturbation of its random motion.
In contrast, a static field
$H_z=620 \, {\rm A/m}$
is able to confine the
particle motion
on a narrow region
between two BWs, i.e. along one ferromagnetic domain, Fig.1(d).
Particles located above the domain of 
opposite magnetization, 
are repelled
from there since,
the energy landscape features 
a local maximum, blue curve in Fig.1(c).\\
In order to characterize
the diffusive properties
along the magnetic conduit,
we measure
the mean squared displacement (MSD)
of the
colloidal particles,
$\langle \Delta y^2 \rangle =\langle [y(t)-y(0)]^2 \rangle \sim t^{\alpha}$,
being $y$ the direction
perpendicular to the magnetic conduit.
Here $\langle ... \rangle$ denotes an average over $\sim 50$
independent trajectories, and
$\alpha$ is
the exponent of the power law which
is used to
distinguish between normal diffusive ($\alpha =1$)
dynamics from the anomalous ones ($\alpha \neq 1$).
In the first case, when
$\langle \Delta y^2 \rangle \sim t$,
one can extract the
effective diffusion coefficient
from the slope of the MSD as $D_y=\lim_{t\to\infty} \, \langle \Delta y^2 \rangle/2t$.
Fig.3(a) shows several MSDs for a $360nm$ particle
in a magnetic conduit 
at different values of the applied field $H_z$.
For $H_z=0$, the motion is isotropic and
the colloidal particles show 
the same diffusion coefficient in the 
2d plane,
$D_x=D_y=0.78 {\rm \mu m^2/s}$
for the $360nm$ particle.

We characterize the zero field dynamics 
for all three types of paramagnetic colloids, 
and provide in Table 1 the comparison between the 
effective diffusion coefficients measured on top of a non-magnetic 
glass substrate and on top of the FGF. The first ones 
(column 2, Table 1) are closer to the theoretical 
predictions obtained from the Stokes-Einstein 
relationship given by $D_x=k_B T/\zeta$. 
Here $\zeta = 3 \pi \eta  f d_p$ is the friction 
coefficient of a particle immersed in water
($\eta = 10^{-3} {\rm Pa \cdot s}$), 
and $f$ a correction factor which takes 
into account the proximity of the wall~\cite{Hap73}. 
For the glass substrate (column 3, Table 1) we take $f =1$ 
since the particles easily detach from the plane 
in absence of any  magnetic attraction. 
In contrast the FGF attracts the magnetic 
particles reducing their effective Brownian
diffusion. 
Here the particles will experience higher 
friction since $f > 1$, and from the expression of $f$,~\cite{Note1} we 
can estimate an average elevation of $80nm$ of the 
nanoparticle surface from polymer film coating the FGF. 
The particle levitation above the FGF results 
from the balance between the magnetic attraction 
and the repulsion arising from electrostatic and 
steric interactions with the polymer film 
coating the FGF.

When applying the field, the particle motion is confined
within the
magnetic conduit, and the MSD rapidly saturates
to a plateau which indicates the
average maximum space the particle can explore.
Assuming that the complex trapping potential 
can be approximated locally by a
parabolic one of type
$U_m(y)\sim\frac{1}{2}k_e y^2$,
being $k_e$ the elastic constant,
one can derive the
corresponding MSD from the
overdamped Langevin equation.~\cite{Uhl39}
For a single magnetic particle in this potential one has
two independent equations of motion,
\begin{eqnarray}
\zeta \frac{dx}{dt} = f_x(t)  \\
\zeta \frac{dy}{dt} = -k_e y +f_y(t) \, \, ,
\label{Langevin}
\end{eqnarray}
where $f_i$ are the random forces ($i=(x,y)$)
with properties, $\langle f_{i} (t) \rangle=0$
and $\langle f_{i} (t) f_{i} (t') \rangle=2\zeta k_B T \delta(t-t')$.
Solving these equations give, respectively
\begin{eqnarray}
\langle \Delta x^2 \rangle = 2\frac{k_B T}{\zeta} t=2Dt  \\
\langle \Delta y^2 \rangle = \frac{k_B T}{k_e} [ 1-\exp{(-2 k_e t/\zeta)}] \, \, .
\label{MSDtrap}
\end{eqnarray}
We use Eq.(4) to fit
all curves showed in Fig.3(a),
by keeping constant $\zeta=5.2 \cdot 10^{-3} pN \cdot s/\mu m$,
and varying $k_e$
as a fitting parameter.
The results are presented in Fig.3(b),
which
shows a linear relationship between
$k_e$ and the width of the magnetic conduit $\lambda_{+}$.
As shown in the small inset in Fig.3(b), 
$\lambda_{+}$ which correspond to the sizes of the majority domains, 
varies linearly 
with the applied field $H_z$.
In order to resolve directly
the parabolic shape of the potential,
we also analyze
the distribution of displacements
when individual particles are trapped in the magnetic conduit.
Fig.3(c) shows as a typical example the case
for an applied
field $H_z=620 {\rm A/m}$.
The normalized potential was obtained 
by assuming that the probability
density $p(y)$, inset Fig.3(c),
follows the Boltzmann distribution,
$p(y)=p(0)\exp{(-U_m/k_B T)}$.
Inverting this
equation gives
the potential
$(U_m(y)-U_m(0))/k_BT=\log{[p(0)/p(y)]}$,
which is presented in Fig.3(c).
Fitting this curve gives the value of $k_e=0.044 {\rm pN / \mu m}$,
which is very close to the value obtained 
in the independent fit in Fig.3(b), $k_e= 0.042 {\rm pN / \mu m}$
for $H_z=620 {\rm A/m}$.
This result also enforces the hypothesis that 
that the complex magnetic potential can be approximated 
locally by a parabolic one.\\
We next explore the effect of increasing the particle
concentration
inside the magnetic conduit,
for particle densities $\rho < 0.6 \mu m^{-1}$,
where $\rho=N/L$ is the linear density and $L$
the length of the channel.
We find that increasing the
potential stiffness via the applied magnetic field,
reduces the particle motion from 2d,
where
the submicrometer particles can pass each other,
to strictly one dimension (1d), Fig.4(a).
In the first case, 
two approaching particles can eventually exchange position, 
and this process occur by sliding laterally along 
the channel width rather than overriding one below the other.
In the second case, the particles form a "single file"
system, i.e. they assemble into a 1d chain of
diffusing submicrometer particles unable to pass each other.
Single file diffusion is a phenomenon 
observed in a broad class of system, from zeolites,~\cite{Kag92}
to ion channels,~\cite{Hod1955}
dipeptide,~\cite{Mus14,Dut15}
and carbon nanotubes.~\cite{Das2010}
While the statistical properties of a single file of particles are well understood
and experimentally demonstrated,~\cite{Wei00,Lut04}
the transition from single-file to 2d diffusion
has been much less accessible,~\cite{Luc12} and was only
recently realized with microscopic particles.~\cite{Sie12}
Here we monitor this transition
by measuring the particle distribution $P(y)$ along the channel width,
Fig.4(b).
In agreement with previous numerical results,~\cite{Luc12}
we observe that as the confinement decreases, the distribution
become broader, and its variance increases.
Also, given the parabolic 1d confinement
of the FGF, the distribution of particles $P(y)$
remain symmetric along the center of the magnetic conduit ($y = 0$). 
However, in contrast to the numerical results 
reported previously,~\cite{Luc12}
beyond the crossover point from $2d$ to $1d$,
when the particles start
to pass each other, we did not observe a bimodal
distribution of the particle displacement.
The particles did not form
a two-chain structure which would cause
two small peaks in the $P(y)$,
as observed in simulation for an idealized system
of repulsive microscopic colloids.
This difference may arise from the
fact that the inter-particle interactions are not
strong but rather the system behaves as
a confined hard sphere gas with a short
repulsive potential
resulting from steric interactions.
We also shown that
the particle distribution $P(x)$ along the channel,
inset of Fig.4(b),
behave as a Gaussian (continuous line),
as predicted theoretically in the past~\cite{Jor92,Jor93}.
By analyzing the particle distribution
in Fig.4(b) we find that the crossover from 2d to 1d
occurs for field amplitude $H_z=870 {\rm A/m}$
which corresponds to a standard deviation 
$\sigma_x=0.53 \mu m$ for the $360 nm$ particles.
Smaller particles $270 nm$
require much stronger field to be
confined into a 1d line, $H_z>1240 {\rm A/m}$,
while larger ones ($540nm$)
are already 
unable to pass each other for field $H_z>530 {\rm A/m}$, inset Fig.4(c).
This behaviour is a direct consequence of the linear 
dependence of the potential stiffness with the applied field, 
Fig.2(b), since smaller particles have larger thermal 
fluctuations and thus require an higher field to be confined into a 1d line.
Moreover, the data 
in Fig.4(c) can be well fitted with an
exponential curve, $H_z=H^0_z \exp{(-d_p/\beta)}$
which allow us to estimate as $d_p=75 {\rm nm}$ the minimum particle size
which would be possible to trap in a 1D single file
for a maximum applied field of $H_z=3 {\rm kA/m}$,
beyond which
the magnetic conduit start deforming.\\
For high concentration of particles,
excluded volume effects
become important and cause
clogging inside the magnetic conduit, 
which causes freezing of the colloidal structure
strongly reducing the particles diffusive motion.
In Fig.5 we report the condensation of 
a large collection
of $270nm$ sized particles on
up- or down-magnetized
domain depending on the direction of the applied field.
Upon application of an
oscillating field 
at an angular frequency $\omega=2 \pi {\rm rad/s}$ and an amplitude $H_z^0= 1370 {\rm A/m}$,
$H_z=H_z^0\sin{(\omega t)}$,
the particles can be periodically translated from
one domain to the next, since the
magnetic energy landscape is inverted during each field period,
changing maxima to minima and viceversa.
As shown in VideoS4 in the Supporting Information,
the particles perform a periodic
motion between
the magnetic domains,
which is essentially
synchronous with the driving field.
Once inside a minima, the particles
do not show any net motion
and their dynamics is frozen within 
the cages created by the nearest particles.
Thus stopping the field at a certain
value allows to solidify the pattern
in its current state,
without loosing
particles in the nearest domains.
Since the BW response to the applied field
can be considered instantaneous
if compared to the particle self-diffusion time,
this capability to condense particles
pattern suggests new route
of printing magnetic ink containing nanoparticles,~\cite{Hel2007}
or can be used for novel magneto-optical~\cite{Joh2004} or biosensing
applications.~\cite{Kle11}\\
In summary, we report a simple and robust method
to confine strongly fluctuating
magnetic submicrometer particles on a plane
and tuning their diffusive motion from 1d to 2d.
The transition between these dimensions can be easily induced via 
application of
a relatively small external field.
Although our study is limited to
particles which can be resolve by optical microscopy,
separate model calculations
(not presented here) suggest
that also smaller particles can be equally
confined in the magnetic conduits
when adjusting their height above
the FGF surface.
Since the complex magnetic potential generated by pairs of
BWs can be well approximated by an idealized
parabolic well,
the present pilot system could serve
as a testbed for fundamental studies 
of transport and diffusion
at the nanoscale.
Finally, the implementation of 
different FGF film with in 
plane magnetization would in principle 
allow manipulation of a single particle within an ensemble 
using magnetic domain wall tips.~\cite{Hlh03}

\begin{table}[ht]
\caption{Effective translational diffusion coefficients at $H_z=0$ above a glass plate ($D_{g}$),
and above the FGF ($D_{FGF}$). The theoretical values are calculated for $f=1$ ($D_{g}^{th}$)
and for a particle elevation of $80 nm$ ($D_{FGF}^{th}$).}
  \begin{tabular}{ccccc}
    \hline
   Particle size ($nm$) &  $D_{g}  ({\rm \mu m^2/s})$  &  $D_{g}^{th} ( {\rm \mu m^2/s})$ & $D_{FGF} ( {\rm \mu m^2/s})$  &  $D_{FGF}^{th} ( {\rm \mu m^2/s})$\\
    \hline
    $270$   & $1.58 \pm 0.05$  & $1.61$ & $1.04 \pm 0.05$   & $1.03$\\
    $360$   & $1.18 \pm 0.03$   & $1.20$ & $0.78 \pm 0.03$   & $0.72$\\
    $540$   & $0.81 \pm 0.02$  & $0.80$  & $0.52 \pm 0.03$   & $0.44$\\
    \hline
  \end{tabular}
\end{table}

%
\begin{figure}[t]
\begin{center}
\includegraphics[width=0.9\columnwidth,keepaspectratio]{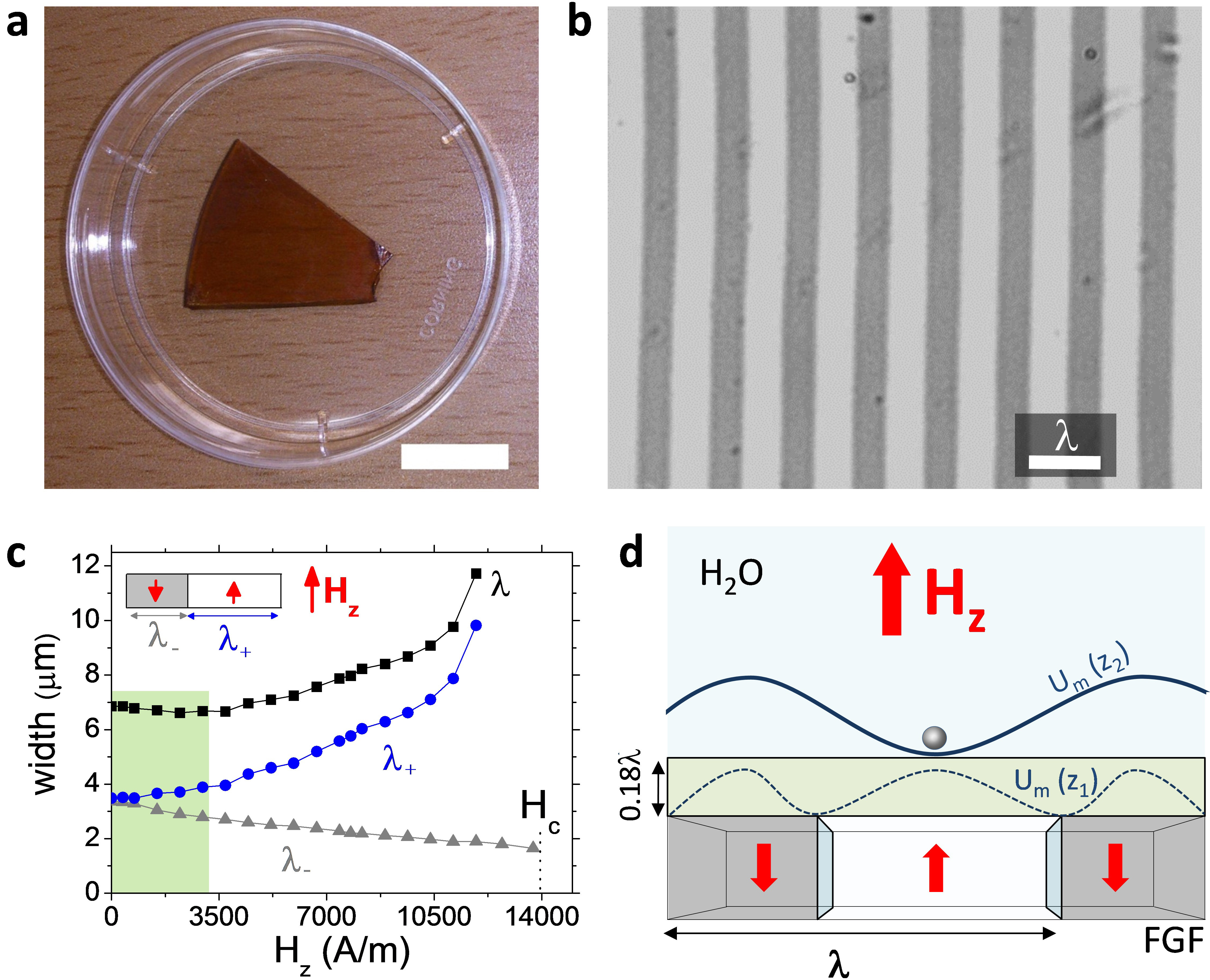}
\caption{
(a) Image of a ferrite garnet film (FGF)
on a gadolinium gallium garnet
substrate, scale bar is $1 cm$.
(b) Polarization microscope image showing a
$57 \times 47 \, \mu m^2$ section of a
magnetic stripe pattern in a FGF film, scale bar is
$\lambda=6.8 \mu m$.
(c) Domain width versus amplitude of the
magnetic field applied perpendicular to the film, $H_z$.
The applied field increases (decreases)
the width of the
domain with parallel (antiparallel) magnetization, $\lambda_{+}$ ($\lambda_{-}$, respectively),
being $\lambda=\lambda_{+}+\lambda_{-}$.
$H_c=13900 \, {\rm A /m}$ denotes the critical field at which
all the antiparallel (grey) domains disappear.
Shaded green area indicates the region where 
the measurements are taken.
(d) Schematic of a FGF film with one magnetic
particle
and subjected to the perpendicular field $H_z$.
Red arrows within the FGF denote the magnetization
direction,
dashed and continuous blue lines
refer to the magnetic potential $U_m$
for a submicrometer particle
at the FGF surface, elevation $z_1$,
and above the photoresist layer at elevation $z_2=z_1+h$,
with $h=1.2 {\rm \mu m}$.}
\label{fig_1}
\end{center}
\end{figure}
%
%
\begin{figure}[t]
\begin{center}
\includegraphics[width=\columnwidth,keepaspectratio]{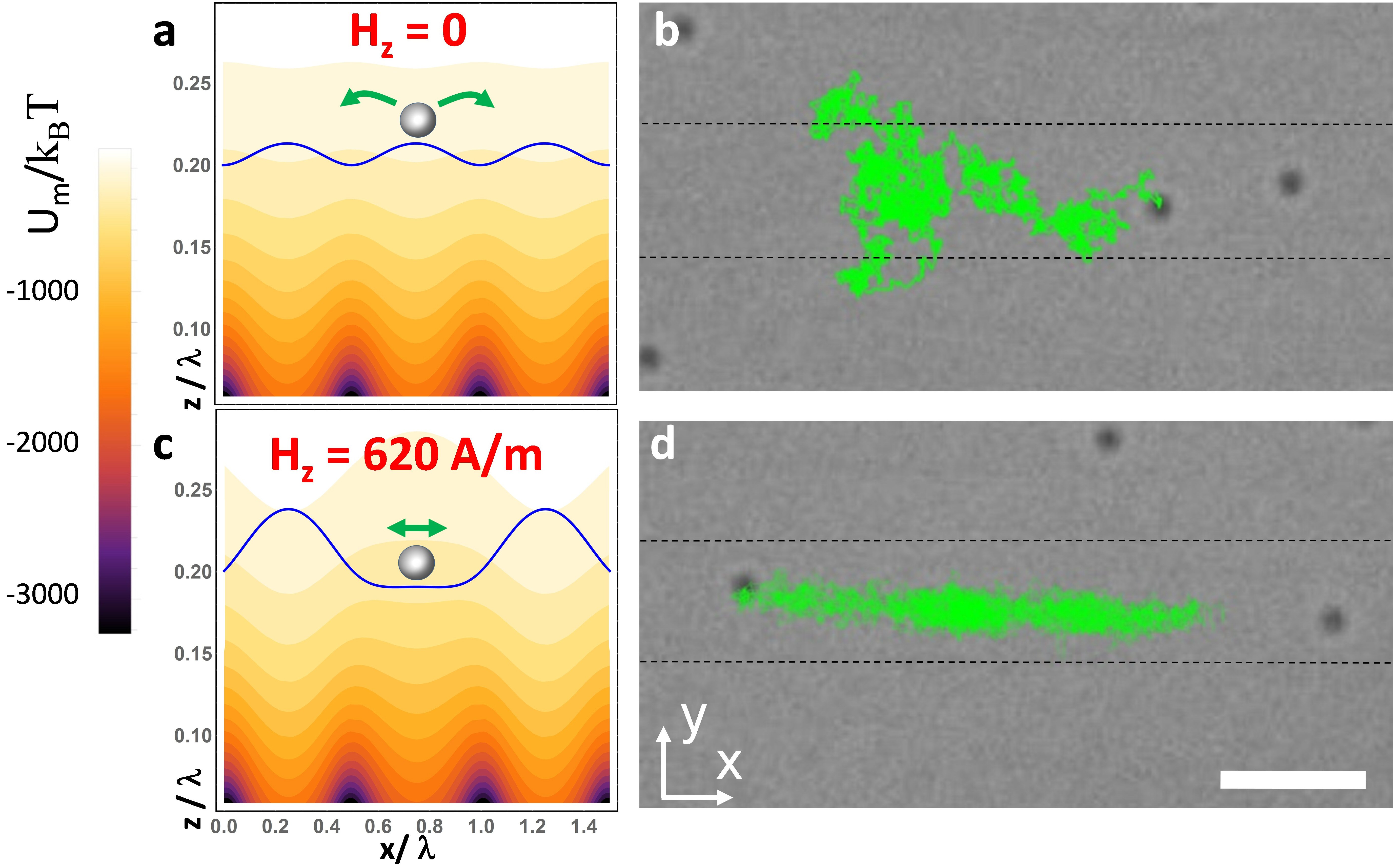}
\caption{(a,c) Color coded
energy landscape $U_m/k_B T$
of one submicrometer particle in absence of field, $H_z=0$ (a), and under
an applied field
$H_z=620 {\rm A/m}$ (c).
Continuous blue lines indicate the
corresponding scaled potential $U_m-\langle U_m \rangle$
plotted at the particle elevation, $z=0.20 \lambda$.
(b,d) Corresponding microscope snapshots of one $360nm$
particle with superimposed
the trajectory (green line)
for $H_z=0$ (b), and $H_z=620 {\rm A/m}$ (d).
Scale bar is $5 \, {\rm \mu m}$,
dashed lines denote the location of the BWs.
See also corresponding MovieS2 in the Supporting Information.}
\label{fig_1}
\end{center}
\end{figure}
%
%
\begin{figure}[t]
\begin{center}
\includegraphics[width=\columnwidth,keepaspectratio]{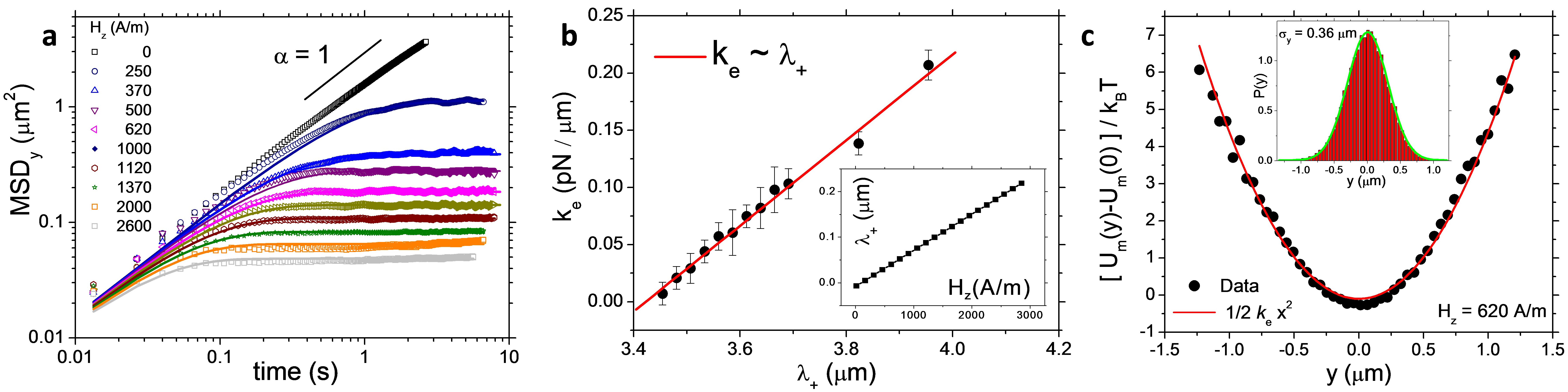}
\caption{(a) Log-log plot of the
mean square displacement (MSD$_y$)
for different applied magnetic fields.
Continuous lines are fit to the data 
using Eq.(4) in the main text.
In absence of field (black squares)
the particle displays normal diffusion
with slope $\alpha=1$.
(b) Spring constant $k_e$ measured
at different width $\lambda_{+}$
of the magnetic channel
for a $360{\rm nm}$ particle. 
Continuous red line denotes linear fit
to the data.
Inset shows the linear relationship between
$\lambda_{+}$ and the applied field $H_z$
for the range of field amplitude used.
(c) Normalized magnetic potential of a $360nm$
particle measured from the particle
fluctuations. The red line represents a parabolic fit
to the data using a stiffness $k_e=0.044 {\rm pN / \mu m}$
for an applied field $H_z=620 {\rm A/m}$.
Small inset at the top shows the
normalized position distribution,
fitted to a Gaussian function (green line) having dispersion 
$\sigma_y=0.36 {\rm \mu m}$.}
\label{fig_1}
\end{center}
\end{figure}
%
%
\begin{figure}[t]
\begin{center}
\includegraphics[width=\columnwidth,keepaspectratio]{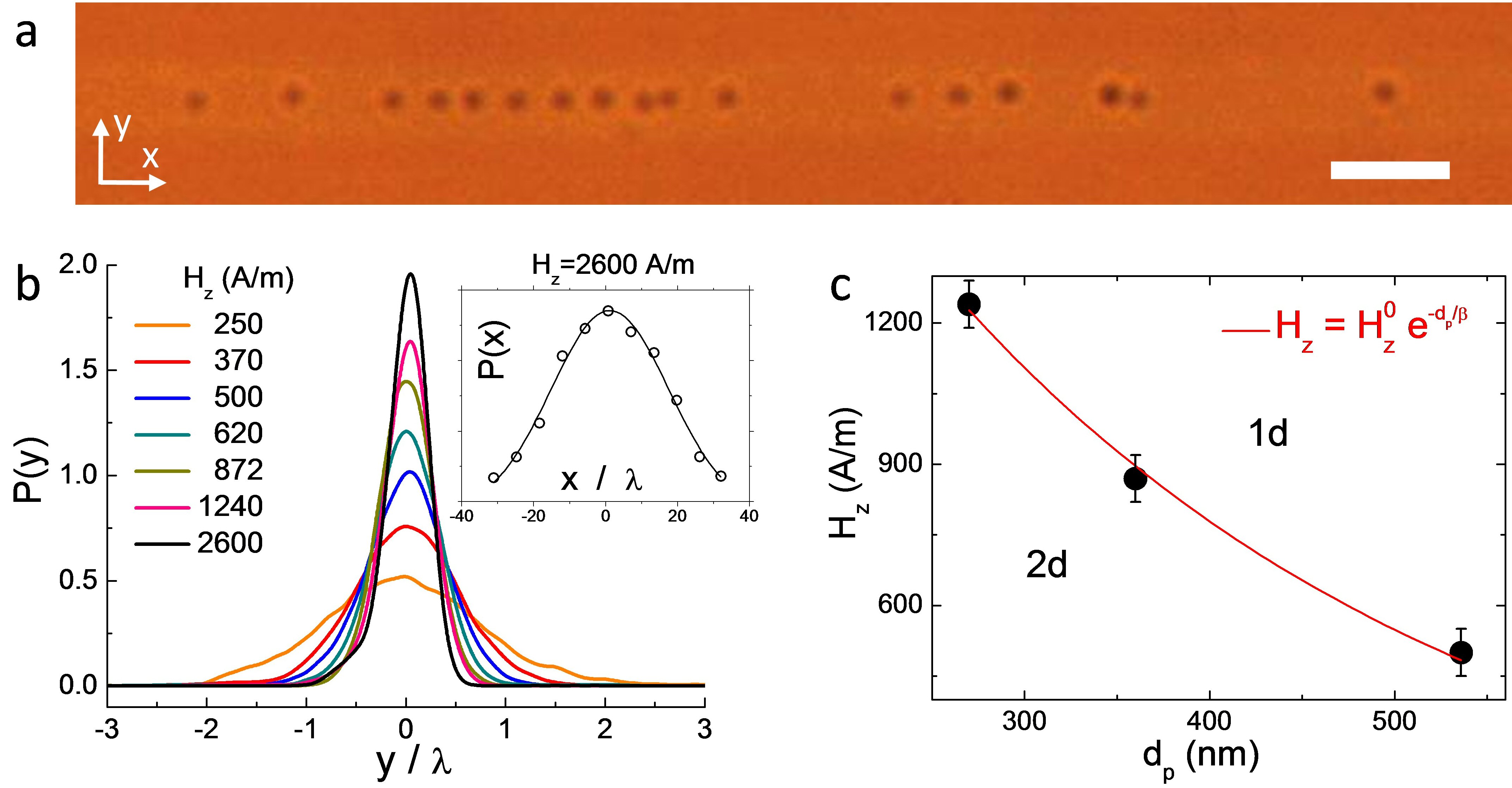}
\caption{(a) Snapshot of a single file composed by $17$
particles ($360nm$) subjected to a field $H_z=1240 {\rm A/m}$. Scale bar is $5 \mu m$.
Corresponding VideoS3
is included in the Supporting Information.
(b) Probability distribution of the
particle position $P(y)$
along the direction perpendicular to
the magnetic conduit
for a linear density $\rho \sim 0.3 {\rm \mu m^{-1}}$.
Inset shows the probability distribution $P(x)$
along the magnetic channel
for an applied field $H_z=2600 {\rm A/m}$,
and with a Gaussian fit (continuous line).
(c) Field dependence of the transition region from
2d to 1d motion versus particle size $d_p$.
The continuous red line is an exponential fit $H_z=H^0_z \exp{(-d_p/\beta)}$
to the data points, with $H_0^z=3158 {\rm A/m}$ and $\beta=285 {\rm nm}$.}
\label{fig_4}
\end{center}
\end{figure}
%
%

\begin{figure}[t]
\begin{center}
\includegraphics[width=\columnwidth,keepaspectratio]{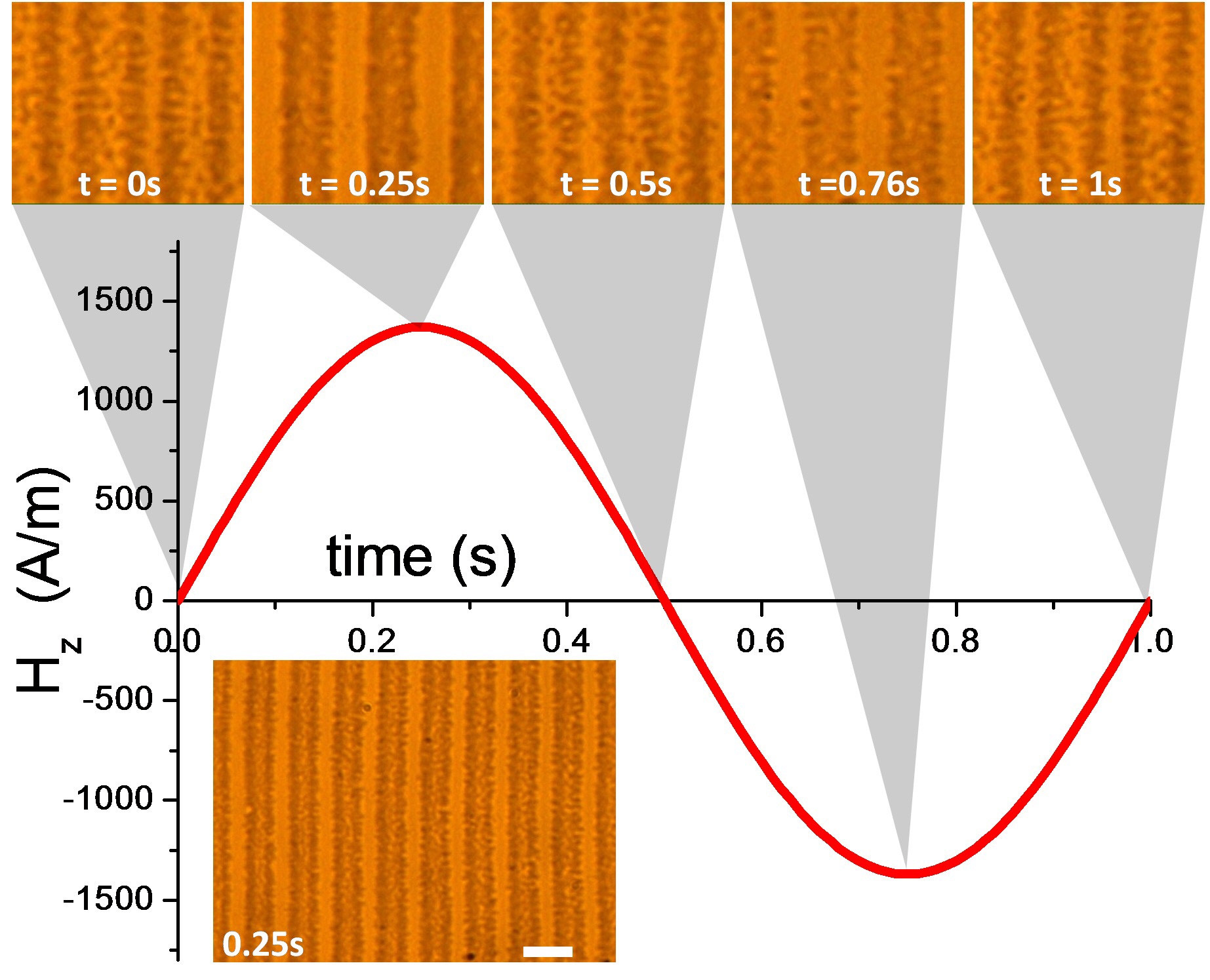}
\caption{Top: series of microscope images of a small section ($16.8 \times 19.2 \mu m^2$)
of an ensemble of $270nm$ particles deposited above 
a FGF film at different value of the applied field.
See VideoS4 in the Supporting Information.
Bottom: external magnetic field $H_z$ oscillating around the $z$
direction with angular frequency $\omega=2 \pi {\rm rad/s}$ and
amplitude $H_z^0= 1370 {\rm A/m}$. Small inset shows
the full overview ($64.3 \times 48.2 \mu m^2$) of the condensed pattern at $t=0.25s$,
scale bar is equal to $\lambda$.
Corresponding VideoS4
is included in the Supporting Information.}
\label{fig_5}
\end{center}
\end{figure}

\begin{acknowledgement}
We acknowledge stimulating discussions with
Arthur Straube.
P. T. acknowledges support from the ERC Starting
Grant ``DynaMO'' (no. 335040), from
the MINECO (Program No.RYC-2011-07605 and
FIS2013-41144-P) and DURSI (2014SGR878).
J. M. S. acknowledges support from
Mineco (FIS2015-66503-C3-3P)
\end{acknowledgement}

\begin{suppinfo}
Four experimental videos (.WMF)
showing the 
dynamics of magnetic colloidal particles
above the FGF film. 
\end{suppinfo}

\bibliography{bibliography}
\end{document}